\documentclass[aps,prd,amsmath,amsfonts,amssymb,eqsecnum,nofootinbib,
  floatfix,secnumarabic,preprintnumbers,superscriptaddress,nobalancelastpage,
  onecolumn,notitlepage]{revtex4-2}
\usepackage{color,graphicx,hyperref,dsfont,slashed,multirow}
\usepackage{placeins}
\usepackage{caption}
\usepackage{subcaption}

\hypersetup{bookmarks=true,unicode=true,pdftoolbar=true,pdfmenubar=true,
  pdffitwindow=false,pdfstartview={FitH},pdftitle={mBLRISS},
  pdfauthor={~A.~Kuday,F.~\"{O}zok, ~U. ~Saka}, pdfsubject={Subject},
  pdfcreator={Creator},pdfproducer={Producer}, pdfkeywords={dark matter}{Effective Field Theory}{Relic Density}{Relic Abundance}{Monojet}{Dijet}{Missing Transverse Energy}
  pdfnewwindow=true,colorlinks=true,
  linkcolor=blue,citecolor=magenta,filecolor=magenta,urlcolor=cyan}
\newcommand{\be}{\begin{equation}}
\newcommand{\ee}{\end{equation}}
\def\bsp#1\esp{\begin{split}#1\end{split}}
\renewcommand{\figureautorefname}{Fig.}

\def\sectionautorefname~#1\null{Sec.~(#1)\null}
\def\subsectionautorefname~#1\null{sub--Sec.~(#1)\null}
\def\figureautorefname~#1\null{Fig.~#1\null}
\def\tableautorefname~#1\null{Table~#1\null}
\def\equationautorefname~#1\null{Eq.~#1\null}

\begin{document}


{\title{Scalar Dark Matter Analysis of 6-Dimensional Effective Field Theory}

\author{Ay\c{s}e  El\c{c}{\.i}bo\u{g}a  Kuday}
\email{akuday@kho.msu.edu.tr}
\affiliation{Department of Fundamental Sciences, Turkish Military Academy, National Defence University,  06654, Ankara, TURKEY}

\author{Ferhat \"Ozok}
\email{Ferhat.Ozok@cern.ch}
\affiliation{Department of Physics, Mimar Sinan Fine Arts University,  34380, {\.I}stanbul, TURKEY}

\author{Erdin\c{c} Ula\c{s} Saka}
\email{ulassaka@istanbul.edu.tr}
\affiliation{{\.I}stanbul University Faculty of Science Department of Physics, 34134 , {\.I}stanbul, TURKEY}

\vspace{10pt}
\begin{abstract}
We give a prescription how high energy physics tools can be used to perform scalar dark matter analysis. We also present the analysis results of scalar dark matter in the context of 6 dimensional  Effective Field Theory.

\end{abstract}

\keywords{scalar dark matter, effective field theory, relic density, direct detection, indirect detection, lhc, fcc, high energy collider, dijet}

\maketitle

\section{Introduction}\label{sec:intro}

Dark matter is a theoretical substance that is responsible for about \%84 of matter compound of universe. Yet, there is no known properties except its non-luminous and non-baryonic characteristics. Up to present, all the knowns about its existence is originated from the large-scale galactic observations. Futhermore, its origin and nature still remain to be brought to the light. For this purpose, many models have been proposed to enlighten the facts to be provided about dark matter. The most common agreement is that it is a  weakly interacting massive particle (WIMP). WIMP is a strong proposal to sort out relic abundance of DM in thermal history of the universe. In that context the interactions of WIMP with SM particles may provide vital information about its particle features.   

\section{Scalar Dark Matter of Effective Field Theory}
Dark matter searches are mainly focused on three different direction of searches; direct detection searches by calculating scattering cross sections of DM with heavy nuclei in a shielded underground laboratories, indirect detection searches by calculating final states of DM annihilation in a galactic halo and collider searches by analysing DM production as missing transverse energy. In order to examine dark matter with these search platforms Effective Field Theory (EFT) offers an exclusive approach in a model-independent way. Basically, it is a practical approach to describe interactions between dark sector and SM with a cut-off scale $\Lambda$ which is considered as a heavy mediator in Simplified Models. For a particular spin of DM, all possible interactions between DM and SM particles can be described via effective dark operators which turn out to be effective interaction Lagrangians. For the particular scope of this study, when we considered dark matter as a scalar field in 6 dimensional EFT. In context of DM EFT spin-0, scalar DM field is assumed to be a real gauge singlet. Tree level interactions between this real scalar DM and SM particles can be taken as the Lagrangian form of the operators given in \cite{ref}. Interactions between SM and this real scalar dark matter field can be categorized as three main outlines: only scalar interactions, vector-scalar interactions, scalar-fermion interactions. Interaction Lagrangians in terms of operator interactions can be listed as: 

		\begin{itemize}
		\item Effective only scalar interactions:
			\begin{eqnarray} \label{eq:phivarphi}
			\mathcal{L}_{\phi 1}=\frac{\alpha_{\varphi 1}}{2 \Lambda^2}(\phi^{\dagger}\phi)^2(\varphi \varphi)
			\end{eqnarray}
			\begin{eqnarray} \label{eq:phivarphi2}
			\mathcal{L}_{\phi 2}=\frac{\alpha_{\varphi 2 }}{\Lambda^2}\partial_{\mu}(\phi^{\dagger}\phi)\partial^{\mu}(\varphi \varphi)
			\end{eqnarray}
	
		\item Effective vector-scalar interactions:
			\begin{eqnarray} \label{eq:phivarphi3}
			\mathcal{L}_{\phi 3}=\frac{\alpha_{\varphi 3 }}{\Lambda^2}(\varphi \varphi) [(D^{\mu}\phi^{\dagger}) (D_{\mu}\phi)]
			\end{eqnarray}
		\end{itemize}

		\begin{itemize}
		\item Effective scalar-fermion interactions:
			\begin{eqnarray} \label{eq:evarphi}
			\mathcal{L}_{e \phi}=\frac{\alpha_{e\varphi}}{\Lambda^2}(\varphi \varphi) (\bar{\ell}e\phi)
			\end{eqnarray}
			\begin{eqnarray} \label{eq:dvarphi}
			\mathcal{L}_{d \phi}=\frac{\alpha_{d\varphi}}{\Lambda^2} (\varphi \varphi) (\bar{q}d\phi)
			\end{eqnarray}
			\begin{eqnarray} \label{eq:uvarphi}
			\mathcal{L}_{u \phi}=\frac{\alpha_{u\varphi}}{\Lambda^2}(\varphi \varphi) (\bar{q}u\phi)
			\end{eqnarray}
		\end{itemize}

where, $\varphi$ is real scalar gauge-singlet DM field in 6 dimensional EFT, $\phi$ is the Higgs field in unitary gauge, $\alpha$'s are dark operators represent interaction coupling constants, $\Lambda$ is cut-off scale of effective theory, $D^{\mu}$ is covariant derivative,  $q$ and $\ell$ are, respectively, left-handed quark and lepton doublet fields, $u$,$d$ and $e$ are right-handed singlet fermion fields. Eq. \ref{eq:phivarphi} and Eq. \ref{eq:phivarphi2} indicate the effective only-scalar interactions, Eq. \ref{eq:phivarphi3} indicates effective scalar-vector interactions, and Eq. \ref{eq:evarphi}, Eq. \ref{eq:dvarphi} and Eq. \ref{eq:uvarphi} indicate effective scalar-fermion interactions of 6-dimensional EFT. Feynman diagrams that correspond to the most dominant annihilation processes are depicted in Fig. \ref{fig:feynman-ska}.

\begin{figure}[h]
    \centering
    \includegraphics{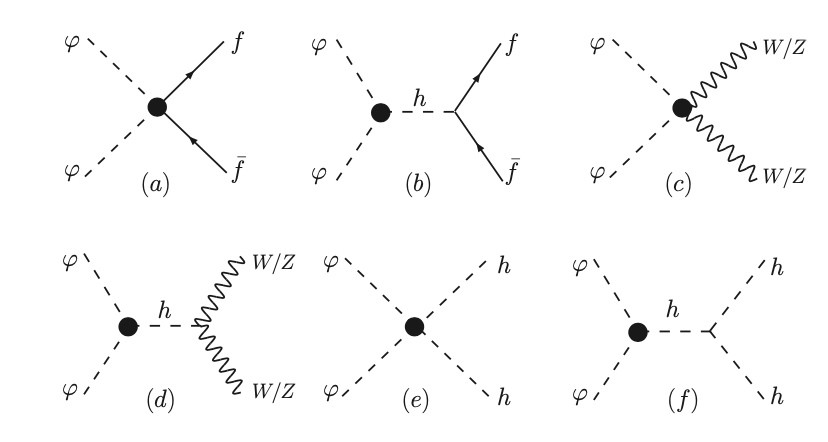}
    \caption{Some Feynman diagrams which contribute to DM annihilation processes dominantly.}
    \label{fig:feynman-ska}
\end{figure}

In Fig. \ref{fig:feynman-ska} the blobs indicate the effective vertices with scalar DM and SM particles. Feynman diagram corresponds to 4-point scalar-fermion interaction is shown Fig. \ref{fig:feynman-ska} (a), Feynman diagrams correspond to 3 point scalar-scalar interactions are shown in \ref{fig:feynman-ska} (b), (d) and (f). 4-point scalar vector and 4-point scalar-scalar interactions are depicted in \ref{fig:feynman-ska} (c) and (e), respectively. These are the most dominant processes that contribute to DM annihilation. Operator formalism of EFT is an obligation in order to construct a model file for Mathematica based Feynrules tool. In an attempt to achieve a well-constructed UFO file of Madgraph/MadDM one can only get all possible combination of DM-SM interactions, obtain decay widths calculated by Feynman rules, check hermicity and charge conservation via using operator formalism. Otherwise, it could be impossible to approach a well-defined model file. In this representation of diagrams and operators we specify a basic frame for scalar candidate. Then, one is left to compute numerical calculation tool by using this framework. 

\section{Status of Scalar Dark Matter of Effective Field Theory}

Any DM model has to be analysed with some specific numerical tool comprehensively. The numerical tool must suffice to compute every vital DM characteristic results, which are put forward by experimental and observational search team. In particular, these constructed numerical tool have to address to calculation of DM relic density, spin-dependent or spin-independent scattering of DM from a heavy nuclei, and DM annihilation cross section in a galactic halo and missing transverse energy (MET) of DM pair production. For the sake of analysis of dark matter along with numerical tool, we generated a model file of scalar DM of 6- dimensional EFT with \textsc{FeynRules} \cite{Feynrules} package to perform a generic analysis of the scalar DM candidate of 6-dimensional EFT. \textsc{FeynRules} package already includes all information about SM of particle physics, we implement all information about interactions given in equations Eq. \ref{eq:phivarphi},Eq. \ref{eq:phivarphi2}, Eq. \ref{eq:phivarphi3}, Eq. \ref{eq:evarphi}, Eq. \ref{eq:uvarphi} and Eq. \ref{eq:dvarphi} and the hermitian conjugates of HC[Eq. \ref{eq:evarphi}, Eq. \ref{eq:uvarphi} and Eq. \ref{eq:dvarphi}]. We also implement the mass of scalar DM as $M_{\varphi}$, coupling parameters as $\alpha$'s, cut-off scale as $\Lambda$, and all other corresponding implicit parameters of in terms of these explicit parameters. The model file is generated to compare with the current experimental and observational data of DM. The model parameters of 6$-$dimensional EFT model of a scalar DM candidate is to be constrained in order to make a reliable analysis.  After generation of the aforementioned  smooth executable UFO model file, \textsc{MadDM} package \cite{Maddm} for DM calculation is used to constrain the model parameters of 6-dimensional EFT for scalar DM candidate. \textsc{MadDM} tool of \textsc{MadGraph} \cite{mg5} has lots of computational skills of dark matter calculations, such as computation of the DM relic density, DM-nucleus scattering cross section for direct detection, and DM-annihilation velocity averaged cross section or  radiation flux of a galactic halo for indirect detection. \\

Any DM candidate is compelled to provide DM relic density limit which is the most well-known characteristic of DM. Therefore, any DM model has to prove that has to ensure the current relic density limits so that in order to determine properties of a DM candidate, one can conclude whether the model is useful or not. In this work in an attempt to constrain the EFT model parameters due to relic density, the relic density is computed for scalar DM candidate of effective theory, and then the results are compared with \textsc{RelicDensity} module of \textsc{MadDM} \cite{Maddm} which is provided from DM relic density data by WMAP ($\Omega h^2=0.11\pm 0.0031$)  \cite{WMAP}.\\

In Fig. \ref{fig:mass_relic_sca} it is showed that how relic density changes according to the mass of scalar DM $M_{\varphi}$ in 6-dimensional EFT. \\

\begin{figure}[h]
    \centering
    \includegraphics{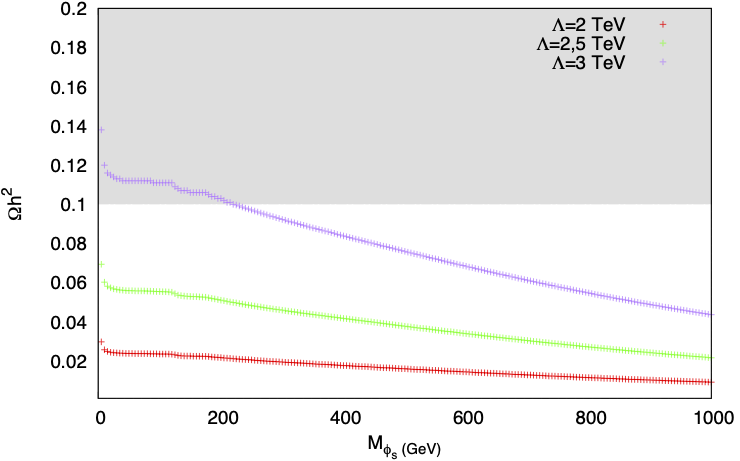}
    \caption{Relic density according to $M_{\varphi}$ mass of scalar dark matter candidate in 6-dimensional EFT.Gray shaded regions are excluded region.}
    \label{fig:mass_relic_sca}
\end{figure}

The shaded regions indicate the excluded region according to current data for thermal relic density of dark matter provided by WMAP. In Fig. \ref{fig:mass_relic_sca}, one can notice that the relic density for scalar dark matter model of 6-dimensional remains proper within the comparison of current WMAP relic density limit. As it can be seen from the Fig. \ref{fig:mass_relic_sca} as cut-off $\Lambda$ increases, the relic density is inclined to be deviate from current upper limit of dark matter relic density. When $\Lambda=3$ TeV, the region $0<M_{\varphi}<225$ GeV does not obey to acceptable limits. \\

When $\Lambda=2$ TeV and $M_{\varphi}=70$ GeV, the model coupling parameter dependence of relic density is plotted in Fig. \ref{fig:alpha_relic_sca}.

\begin{figure}[h]
    \centering
    \includegraphics{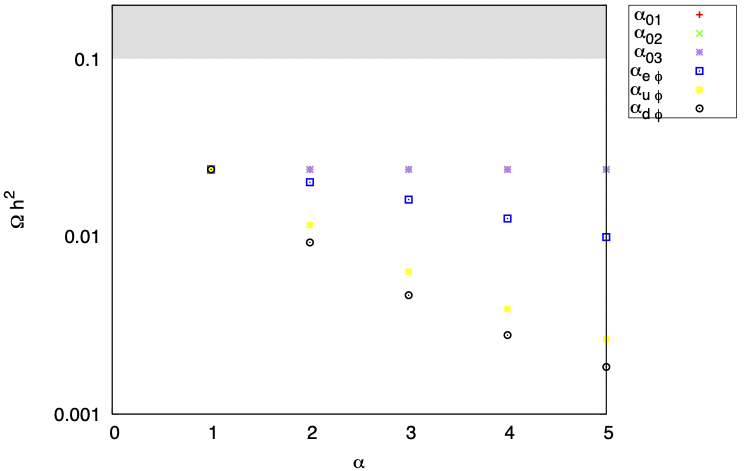}
    \caption{Relic density according to $\alpha$ model parameters of scalar dark matter candidate in 6-dimensional EFT.}
    \label{fig:alpha_relic_sca}
\end{figure}

Fig. \ref{fig:alpha_relic_sca} shows how relic density responses to $\alpha$ model parameters, mainly, how relic density responses to every interaction of scalar dark matter. In the Fig. \ref{fig:alpha_relic_sca} gray shaded region still demonstrates the excluded region of provided WMAP data \cite{WMAP} for DM relic density. It can be seen that most critical deviation of relic density due to the coupling constant stems from the fermionic interactions of scalar DM.  It can be deduced that Eq. \ref{eq:dvarphi} and Eq. \ref{eq:uvarphi} play a big role for generation of relic density on account of altering the relic density of scalar DM relatively larger than other coupling parameters.\\

After constraining model parameters by means of DM relic density, we moved concentration of our study into another DM search area, called indirect detection, which enables to calculate velocity averaged cross section of DM annihilation to SM particles, around a galactic halo. The calculations of velocity avaraged DM annihilation cross sections to SM particle around a galactic halo necessiate \textsc{IndirectDetection} module of \textsc{MadDM} \cite{Maddm}. The scalar DM annihilation velocity averaged cross section to SM particles are shown in Fig. \ref{fig:direct_1500_sca}, Fig. \ref{fig:direct_2000_sca}, Fig. \ref{fig:direct_2500_sca}. \\

\begin{figure}[h]
    \centering
    \includegraphics[scale=0.7]{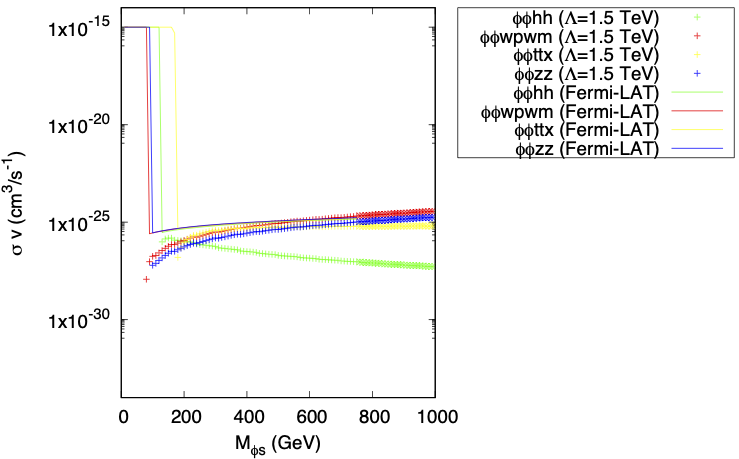}
    \caption{Velocity averaged cross section of scalar Dark Matter annihilation to SM particles, when $\Lambda=1.5$ TeV and $\alpha_{all}=1$. Straight lines demonstrates the upper limits of Fermi-LAT.}
    \label{fig:direct_1500_sca}
\end{figure}

In Fig. \ref{fig:direct_1500_sca} velocity averaged cross section of scalar Dark Matter annihilation to SM particles are shown when $\Lambda=1.5$ TeV and $\alpha_{all}=1$. Straight lines demonstrate the upper limits provided by Fermi-LAT exclusion bounds of velocity averaged DM annihilation to SM particles cross sections for the combined set of dSphs, and each data point stands for the result of 6-dimensional EFT for scalar DM annihilation into SM particles. Jagged data points are unacceptable values originated from validity condition ($Q_{tr}\geq 2 m_{DM}$) \cite{Busoni}. Because annihilation processes related to leptons and other fermions with low mass gives tiny annihilation cross sections, we did not include these results in Fig. \ref{fig:direct_1500_sca}. For the model of scalar DM of 6-dimensional EFT, the masses greater than $M_{\varphi}\geq 620$ GeV is left out in consequence of scalar DM annihilation to $W^+W^-$ bosons. The most annihilation cross sections are obtained by DM pair annihilation to $W^+W^-$, $t\bar{t}$ and $ZZ$ pairs.\\

\begin{figure}[htp]

\begin{subfigure}{\textwidth}
\includegraphics{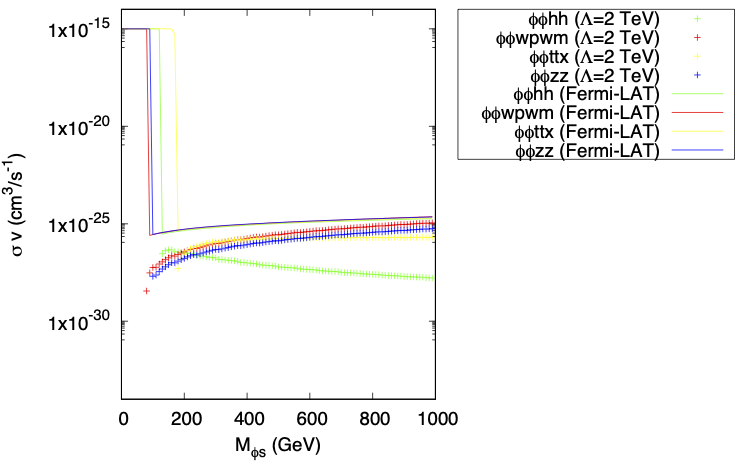}
\caption{Velocity averaged cross section of scalar Dark Matter annihilation to SM particles, when $\Lambda=2$ TeV and $\alpha_{all}=1$. Straight lines demonstrates the upper limits of Fermi-LAT.}
\label{fig:direct_2000_sca}
\end{subfigure}

\bigskip

\begin{subfigure}{\textwidth}
\includegraphics{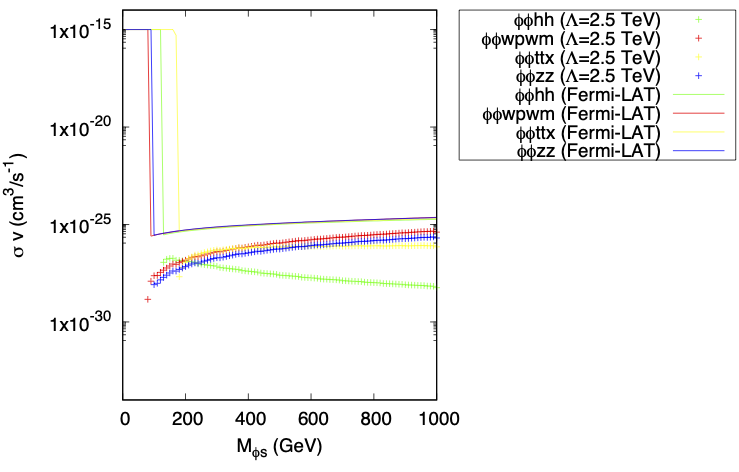}
\caption{Velocity averaged cross section of scalar Dark Matter annihilation to SM particles, when $\Lambda=2.5$ TeV and $\alpha_{all}=1$. Straight lines demonstrates the upper limits of Fermi-LAT.}
\label{fig:direct_2500_sca}
\end{subfigure}

\caption{Velocity averaged cross section of scalar Dark Matter annihilation to SM particles}

\end{figure}

In Fig. \ref{fig:direct_2000_sca} and Fig. \ref{fig:direct_2500_sca} velocity averaged cross section of scalar Dark Matter annihilation to SM particles are shown. Straight lines still demonstrate the upper limits provided by Fermi-LAT exclusion bounds of velocity avaraged DM anniliation to SM particles cross sections for the combined set of dSphs, and each data point is obtained from the result of 6-dimensonal EFT for scalar DM anniliation into SM particles according to cut-off scales $\Lambda=2$ TeV and $\Lambda=2.5$ TeV. One can easily figure out that when $\Lambda=2$ TeV and $\Lambda=2.5$ TeV and $\alpha_{all}=1$ the scalar DM EFT model is consistent with annihilation to SM particles supplied by Fermi-LAT data for mass range $0<M_{\varphi}<1$ TeV. Major limitation by comparing mass from indirect detection with $\alpha_{all}=1$ only comes from the case of  annihilation to W-boson pair when cut off scale is $\Lambda=1.5$ TeV.\\

Lastly, we take our research to collider searches, where DM pair production come along with SM particles. In 6$-$dimensional EFT, the most significant contribution to pair production of scalar DM is given with the process of scalar DM pair production with dijet process. In collider searches, DM is searched as a missing transverse energy (MET). In collider searches of DM, the most dominant process is missing transverse energy ($\slashed{E}_T$) with dijet.  $\varphi\varphi_{jj}$ final state at LHC with $13$ TeV center of mass energy has been examined by scanning different values of cut$-$off $\Lambda$ and $\alpha=1$ values of model coupling parameters.  For DM production with dijet, the model file generated with \textsc{FeynRules} \cite{Feynrules} was included in to \textsc{MG5\_aMC\_v2\_6 \_2} model folder. According the limitation from DM relic density and velicity averaged cross sections of DM  annilihation in indirect detection, $M_{\varphi}=500$ is GeV is acceptable for three specific cut-off scale values, $\Lambda=1.5$ TeV, $\Lambda=2$ TeV and $\Lambda=2.5$ TeV. Concerning the processes of DM pair production at LHC, when the mass $M_{\varphi}=500$ GeV and $\alpha_{all}=1$, $50000$ number of events is generated via MG5 by using cuts $-2.5<\eta<2.5$, transverse momentum of leading jet $p_T\geq 120$, and $\slashed{E}_T\geq250$ GeV. Resulting cross sections at 13 TeV LHC for these processes are given in Table \ref{tab:sca-dijet}.

\begin{table}[h]
\begin{tabular}{|l|l|l|l|l|}
\hline
                                                    & $\Lambda=1.5$ TeV     & $\Lambda=2$ TeV      & $\Lambda=2.5$ TeV    & $\Lambda=3$ TeV      \\ \hline
$\sigma\mid_{(pp\rightarrow\varphi\varphi j j)}$ pb & $3.03 \times 10^{-8}$ & $3.02\times 10^{-9}$ & $2.5\times 10^{-10}$ & $1.1\times 10^{-10}$ \\ \hline
\end{tabular}
\caption{The cross section for scalar DM pair + dijet process at LHC with 13 TeV center of mass energy, for 50000 number of events and with $-2.5<\eta<2.5$, $p_T>120$ GeV ve $\slashed{E}_T>250$ GeV kinematical cuts.}
\label{tab:sca-dijet}
\end{table}
In Table \ref{tab:sca-dijet} the results were obtained by \textsc{Pythia6} shower/hadronization, \textsc{Delphes} detector simulation modules of \textsc{MG5\_aMC\_v2\_6\_2} and default pdf sets \textsc{nn23lo1} . At this point we have emphasize that, obtained missing $\slashed{E}_t$ with dijet cross sections are too small to produce any kinematic distribution of $\eta$, $p_{T,j}$ and $\slashed{E}_T$. Nevertheless, the background cross section with same cuts reads remarkably higher value than the signal cross section.

\begin{table}[h]
\begin{tabular}{|l|l|}
\hline
$\sigma_{pp>jj+\slashed{E}_T}$ (pb) (background) & 
$ 2.1 \times 10^{-3}$  \\ \hline
\end{tabular}
\caption{Background cross section for $\slashed{E}_t + dijet$ process at LHC with 13 TeV center of mass energy, for 50000 number of events and with $-2.5<\eta<2.5$, $p_T>120$ GeV ve $\slashed{E}_T>250$ GeV kinematic cuts.}
\label{tab:dijet-bg-50k}
\end{table}

   By searching different values of model parameters at 14 TeV LHC, the signal cross sections of the  missing $\slashed{E}_T$ with dijet process seem too weak compared with the background of same process. Due to suffering from the small cross section of this process, it is not possible to generate any event. We can conclude that it is not convenient to search scalar DM production with dijet at hadron colliders because of the negligible contributions to DM relic.
\section{Summary and Conclusion}

In this paper we present an investigation of 6-Dimensional  Effective Field Theory of DM where it is considered as a real scalar gauge-singlet DM field. Aim of this work is to construct numerical tools in order to analyse this DM candidate in different search areas. The main analyses of this work can be categorised in three different search class; calculation of DM relic density, spin-dependent or spin-independent scattering of DM from a heavy nuclei within direct search experiments, and DM annihilation cross section in a galactic halo the within indirect search observations, and missing transverse energy ($\slashed{E}_T$) of DM pair production with dijet within colliders. Both DM relic density and velocity averaged annihilation cross section of indirect searches are able to limit large parameter range of 6-dimensional EFT for this scalar DM candidate. For this scope, at first, thermal relic density of scalar dark matter in 6-dimensional EFT is calculated for wide range of each parameters of the effective model and obtained results of relic density for this scalar candidate is compared with the result of current relic density of dark matter provided with WMAP data. By scanning cut-off scale, coupling constant of effective theory and mass of DM candidate, we managed to compare the relic density results that are responsible for the effective theory with the current value of WMAP data for thermal relic density of DM. The validity condition is maintained in all calculations. As it is seen clearly from the Fig. \ref{fig:mass_relic_sca} for values of $\Lambda=2$ TeV and $\Lambda=2.5$ TeV, the relic density is compatible with the current relic density limit of WMAP data \cite{WMAP}. As cut-scale $\Lambda$ values increase relic density may excess the upper limit provided by WMAP. For scalar dark matter of 6$-$dimensional EFT, when cut-off scale is taken $\Lambda=3$ TeV, for the range of DM mass, masses less than $0<M_{\varphi}<225$ GeV, thermal relic density of the scalar DM excess upper limit of DM relic density which is provided by WMAP data. So, it can be concluded that the mass range between $0<M_{\varphi}<225$ GeV is not acceptable. Later, we examine how coupling parameters of the model change the thermal relic density. As it is displayed from the Fig. \ref{fig:alpha_relic_sca} the most significant contribution comes from the interactions between scalar DM and fermions. So interaction of fermion interactions in the Eq. \ref{eq:dvarphi} and Eq. \ref{eq:uvarphi} are the most dominant interactions when generating thermal relic density. Then, by continuing with another DM search area, for figuring out spin-dependent or spin-independent scattering of DM from a heavy nuclei direct detection, results of DM-nucleus interactions are tried to be computed and compared with the results of the direct detection experiment, LUX, PICO60 and XENON1T. Any  transition amplitude are not been able to be produced. Laterly, we give our results of velocity averaged cross section of DM annihilation to SM particles, around a galactic halo. In  Fig. \ref{fig:direct_1500_sca} when $\Lambda=1.5$ TeV and $\alpha_{all}=1$ the most dominant processes of annihilation of scalar DM to SM particle is shown. Most velocity averaged cross sections of the processes of DM annihilation into SM particles stay in proper limit according to cross sections of DM pair annihilation to SM particle provided by Fermi-LAT data, in Dwarf
spheroidal galaxies (dSPhs). DM pair annihilation to $W^+W^-$ is not admissible when the mass of scalar dark matter in the range $M_\phi\geq620$ GeV, because this interaction cross section exceeds Fermi-LAT data. However, Fig. \ref{fig:direct_2000_sca} and \ref{fig:direct_2500_sca} show that there is no improper mass region for scalar dark matter, exceeds current upper limit of Fermi-LAT, DM annihilates to SM particles. Lastly, dark matter production at 14 TeV LHC is investigated. For DM pair production processes at 14 TeV LHC, for processes of accompanying dijet to missing transverse energy ($\slashed{E}_T$), obtained cross sections are listed in Table \ref{tab:sca-dijet}. Cross section of DM pair production with dijet processes of possible background missing $\slashed{E}_T$ with dijet is listed in Table \ref{tab:dijet-bg-50k}. Background cross section is fairly large than signal. Also, we can conclude that the cross section of signal is too small to generate any kinematical distribution.


 


\end{document}